# Processing of Photonic Crystal Nanocavity for Quantum Information in Diamond

*Igal Bayn[a,*], Boris Meyler[a], Alex Lahav[a], Joseph Salzman[a], Rafi Kalish[b], Barbara A. Fairchild[c], Steven Prawer[c], Michael Barth[d], Oliver Benson[d], Thomas Wolf[e], Petr Siyushev[e], Fedor Jelezko[e] and Jorg Wrachtrup[e]*

[a] Department of Electrical Engineering and Microelectronics Research Center, Technion Haifa, 32000, Israel

[b] Department of Physics and Solid State Institute, Technion Haifa, 32000, Israel

[c] School of Physics, The University of Melbourne, Melbourne, Australia, 3010

[d] Nano-Optics, Institute of Physics, Humboldt-University Berlin, Hausvogteiplatz 5-7, D-10117 Berlin, Germany

[e] Physical Institute, University of Stuttgart, Stuttgart 70550, Germany

---

[*]Corresponding author. E-mail address: eebayn@techunix.technion.ac.il (Igal Bayn)

**Abstract**

The realization of photonic crystals (PC) in diamond is of major importance for the entire field of spintronics based on fluorescent centers in diamond. The processing steps for the case of diamond differ from those commonly used, due to the extreme chemical and mechanical properties of this material. The present work summarizes the state of the art in the realization of PC's in diamond. It is based on the creation of a free standing diamond membrane into which the desired nano-sized patterns are milled by the use of Focused-Ion-Beam (FIB). The optimal fabrication-oriented structure parameters are predicted by simulations. The milling strategies, the method of formation the diamond membrane, recipes for dielectric material-manipulation in FIB and optical characterization constraints are discussed in conjunction with their implication on PC cavity design. The thus produced structures are characterized via confocal photoluminescence.

# 1. Introduction

Quantum Information Technology (QIT) may revolutionize many photonic areas, amongst others computing and communication. QIT may offer solutions to problems that nowadays computers are inefficient in, such as certain hidden subgroup problems (e.g. factorization), ultimately secure communication, and modeling of many-body quantum systems [1]. Diamond is a material which holds much promise for such applications as it hosts atomic defects of very suitable luminescence and quantum properties, such as the negatively charged Nitrogen-Vacancy (NV⁻) center. A platform for QIT design and realization with potential implementation by a technology suitable for diamond that does not interfere with that of silicon fabrication is highly desirable. In addition, since the scale of quantum bit (qubit) integration will ultimately define the computational power of a quantum computer and the most successful experience in the large scale integration comes from microelectronics, naturally, the implementation of QIT in diamond will benefit if it relies on a similar technological background [2]. Here, we describe a diamond photonic approach that potentially satisfies these QIT requirements with emphasis on studying the parameter space for its realization.

The negatively charged nitrogen-vacancy color center (NV⁻) in a single crystal diamond is a novel solid state platform for QIT [3],[4]. The center is formed by the substitution of one carbon atom in the diamond lattice with nitrogen and a neighboring vacancy (see Fig. 1a). The NV⁻ ground ($^3A$) and excited ($^3E$) states are spin triplets (see Fig. 1b). The ground spin ($^3A$) can be read-out optically, manipulated by microwave and spin polarized to the $m_s=0$ state at room temperature. Weak spin-orbit interaction and a spin free neighborhood of the center are required to produce the extremely long life time ($T_1$) and coherence time ($T_2$) of $^3A$ states needed for qubit operation. The NV⁻ center in diamond has been shown to exhibit such long coherence times [3], hence it is an ideal candidate for qubit realization, where the $m_s=0$ and $m_s=-1$ (or $m_s=+1$) form the two level system capable for long-time storage, initialization and manipulation. This system potentially satisfies the hardware conditions mentioned above.

Although single qubit and two-qubit operations based on the NV center in diamond were demonstrated by several groups [5]-[9], the realization of large scale quantum computing requires a controllable interaction between distant qubits which cannot relay on the dipole-dipole mechanism (restricted by a *~10nm* maximal distance between the centers). Therefore, the interaction between distant qubits in a multi-qubit operation should be carried out by photons. One of the attractive ways to implement this is by a photonic crystal (PC) architecture (see Fig. 2a), where each qubit is being registered within a high-*Q* cavity, and the cavities are interconnected by PC waveguides (see figure 2b) [10]. Such a system is an ideal building block for a photonic module [11], which provides a natural method for generating a three-dimensional cluster state based quantum computer architecture [12].

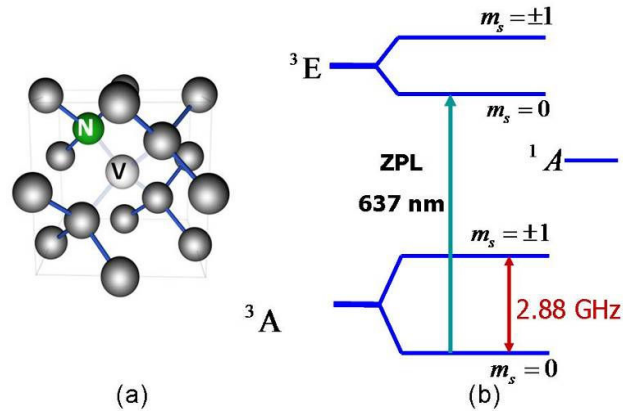

Fig. 1 − Negative-Nitrogen-Vacancy Color Center (NV-). (a) The model of the structure. (b) The energy level scheme for the NV-

The qubit is composed of a single NV- [3],[4],[10] or two NV- [13] coupled to the cavity. The single qubit data is manipulated by application of a microwave pulse (*2.8GHz*) resonant to the

$^3A(m_s=0) \leftrightarrow {}^3A(m_s=\pm1)$ transitions (see Fig. 1b). The qubit initialization to the ground state is produced by broad-band excitation [2]. The waveguide-cavity coupling could be governed by tuning the cavity resonance using the Stark effect [14]. In this way, the requirements for large scale entanglement, single qubit readout and quantum data storage may be met (see Fig. 2b). In addition, the interaction distance which is limited by the waveguide losses can, in principle reach hundreds of microns. The two qubit entanglement [15],[14] and two qubit CNOT [11] protocols based on a Λ-type system where one transition of the Λ is coupled to the cavity mode and another is weakly excited are applicable here. These operations are based on the cavity-cavity or cavity-waveguide-cavity coupling. Combining these protocols, quantum register design [10], and multi-cavity architecture interconnected by the waveguide network produces a core operation packet for quantum computer realization.

In addition to the long-term goal of quantum computing, the implementation of a cavity coupled to a single NV⁻ center in diamond can demonstrate a variety of interesting phenomena. In the strong coupling regime: $g \gg \kappa, \gamma_\perp$, (where $g$ is the Rabi frequency, $\kappa$ is the cavity mode decay rate and $\gamma_\perp$ is the NV⁻ dipole decay rate), dressed states and Rabi oscillations of the optical transition can be studied. Since $\kappa$ is related to the cavity $Q$ ($\kappa=\omega_0/(4\pi Q)$), the $g$ and $\gamma_\perp$ are related via $g = \gamma_\perp (V_0/V_m)^{1/2}$, where $V_0 = (c\lambda^2)/(8\pi\gamma_\perp)$ and $V_m$ is a mode volume [16], we obtain that a value of $Q \sim 3\times 10^3$ is sufficient to achieve strong coupling. In addition, even weak coupling of an NV⁻ to a low-$Q$ cavity, if realized, will allow fluorescence intensity measurement to detect single-spin based on the dim period of the $m_s=1$ state. This measurement is still very difficult due to extremely low photon collection efficiency (~1%) [17]. The realization of cavity-NV⁻ coupling will improve this, since all the photons are emitted into a single cavity mode and the lifetime is modified by the Purcell effect [18].

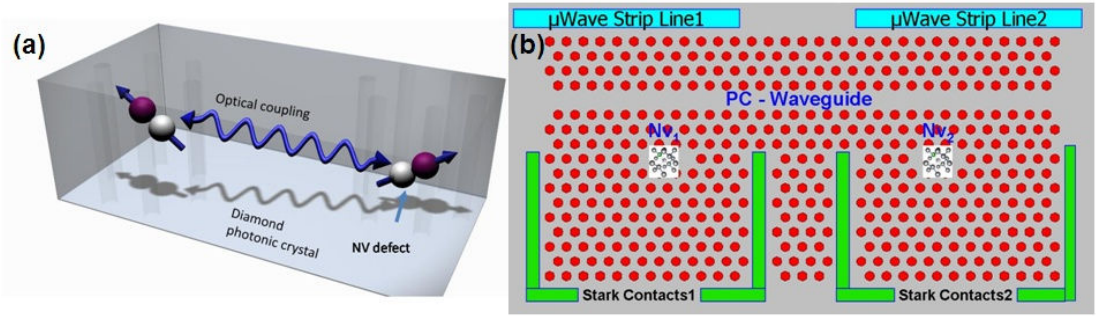

Fig. 2 − (a) Inter-Qubit photon assisted entanglement (b) Two Qubits interconnected by the PC waveguide and manipulated via micro-wave and Stark shift controls [3],[10].

In this paper we present the design considerations and a sequence of processing steps towards fabrication of a diamond photonic crystal structure for QIT. The PC considered here is composed of a diamond membrane suspended in air and periodically modulated by a triangular array of holes. In recent years, significant progress has been made in the design of diamond PC [19]-[22]: an ultra-high-Q cavity design [24], the first nano-crystal-diamond cavity [23] and single $NV^-$ in nanocrystal diamond coupled to a gallium phosphide photonic crystal cavity has been reported [25]-[27]. The implementation of a nano-cavity for QIT in single crystal diamond may exhibit improved optical properties than those of nano-diamond [23]. However, the extreme chemical inertness of single crystal diamond, and the lack of a microfabrication technology, prevented, until now, the implementation of PC in single crystal diamond. Early attempts to realize PC cavities on single crystal diamond were already reported by us and by others [28]-[30], however with no optical characterization. In the present work, the manufacturing process of a 2D-PC slab nanocavity on a single-crystalline diamond membrane is described. A complete set of PC fabrication issues including the design considerations, membrane formation, FIB patterning, and the device relocation technique are discussed.

The fabrication processes include: (1) ion implantation for selected depths graphitization by annealing, (2) graphite removal and etching, (3) patterning and (4) membrane relocation (steps (3) and (4) are performed in situ in the FIB machine (F.E.I. Strata 400 STEM Dual Beam system)). Type-Ib <100> cut single crystal diamond (Sumitomo) was used. The high nitrogen doping level of the Ib diamond results in a high concentration of NV centers, producing a broad photoluminescence spectrum *(570÷800nm)* which allows the confocal micro-photoluminescence spectroscopy-characterization .In order to optimize the performance of the device it is important to explore the parameter space in the design and processing of the photonic devices. We are not aware of any previous report of such work on mono-crystalline diamond for the realization of photonic crystals.

## 2. Photonic Crystal Cavity Design

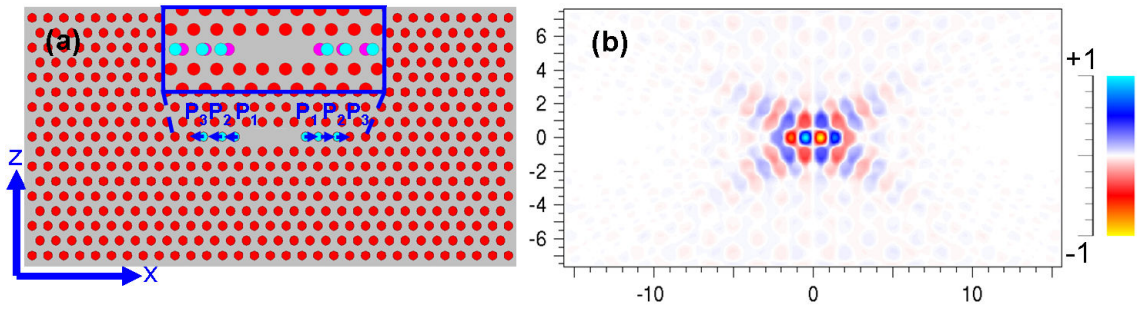

Fig. 3 − The modified 3 missing hole cavity (a) The cavity design. In the inset the shifts of central holes (cyan) are shown in relation to their original positions (magenta) (b) The cavity mode magnetic field $H_y$ at *y=0* (length unit is *a*).

The realization of ultra-high-*Q* of a diamond PC cavity [24] is complicated by the fact that the *Q*-value is sensitive to the fabrication imperfections, which can be of the same order of magnitude as native variations in the material, hence it is unsuitable for the preliminary studies described here. Therefore, we have chosen here to deal with a moderate-*Q* cavity design of the modified 3-missing hole geometry [34],[19]. This cavity is created by missing three central holes of the PC (which are not milled) and a lateral shift of the three holes adjacent to the cavity on both the left and the right sides. The shift distances are denoted by $P_1, P_2, P_3$ (see Fig.3).

The cavity parameters are described in Table-1 in terms of the PC lattice constant (*a*). The choice of *a* defines the entire geometry and the cavity mode wavelength. A Finite-Difference-Time-Domain (FDTD) calculation predicts a quality factor of *Q=8,365* and a mode volume of $V_m=0.921\times(\lambda/n)^3$. Ideally the mode wavelength should be matched to the NV⁻ zero-phonon-line ($\lambda_0=637nm$) and the NV center should be positioned in the electric energy density maximum. In this way the cavity-NV coupling is maximized, which reflects in the highest Purcell coefficient and satisfies the strong coupling conditions. However, due to abundance of NV centers in our sample and their ill defined position within the cavity the realization of a strong coupling regime is impossible. Moreover, since the fabrication of PC includes many uncertainties, such as membrane thickness and hole radius non-uniformity, all of which can potentially modify the mode wavelength, it was chosen here to design the cavity mode wavelength to coincide with the mid-range of Type-Ib diamond photoluminescence (*~700nm*). In this way, even if the geometry is slightly changed during the processing, the cavity can still be characterized by the micro-photoluminescence set-up. Therefore, *a=234nm* is chosen, which corresponds to the basic mode wavelength of $\lambda_0=699nm$.

Table 1 − Modified three missing hole cavity parameters

| *Lattice Constant* | *Thickness T* | *Hole radius r* | *P1* | *P2* | *P3* | *λ* |
|---|---|---|---|---|---|---|
| *a* | *0.96a* | *0.275a* | *0.21a* | *0.025a* | *0.2a* | *2.985a* |
| *234nm* | *225nm* | *64nm* | *49nm* | *6nm* | *46nm* | *699nm* |

## 3. Design and Fabrication Issues

The fabrication tolerances, practical fabrication limitations and their implications on the PC design are:

1) PC hole profile: Usual modeling for cavity calculations assumes vertical cylindrical holes. The holes milled by the FIB depart from the ideal ones, being usually cone shaped (narrowed towards the membrane bottom). Computations show that in the conic profile case, an increase in the inclination angle of the hole side-walls, results in (i) a decrease in the cavity *Q-factor* (Fig. 4(a)), (ii) an increase in mode-volume (Fig.4(b)), (iii) a shift in the eigen-frequency (Fig.4 (c)., and (iv) a drastic decrease in Purcell factor ($F=(3/4\pi^2) \times Q/V_m$) (Fig.4 (d)). For example, a ~4.5° hole wall inclination (common in FIB milling) results in ~24nm shift in the cavity mode wavelength, a deterioration in the *Q* by factor of ~5, an increase in the mode volume ($V_m$) by a factor of ~1.5, and decrease in Purcell coefficient by ~7.7. Therefore, hole milling strategy with a maximum emphasis on the hole verticality (as can be produced by overmilling) is essential.

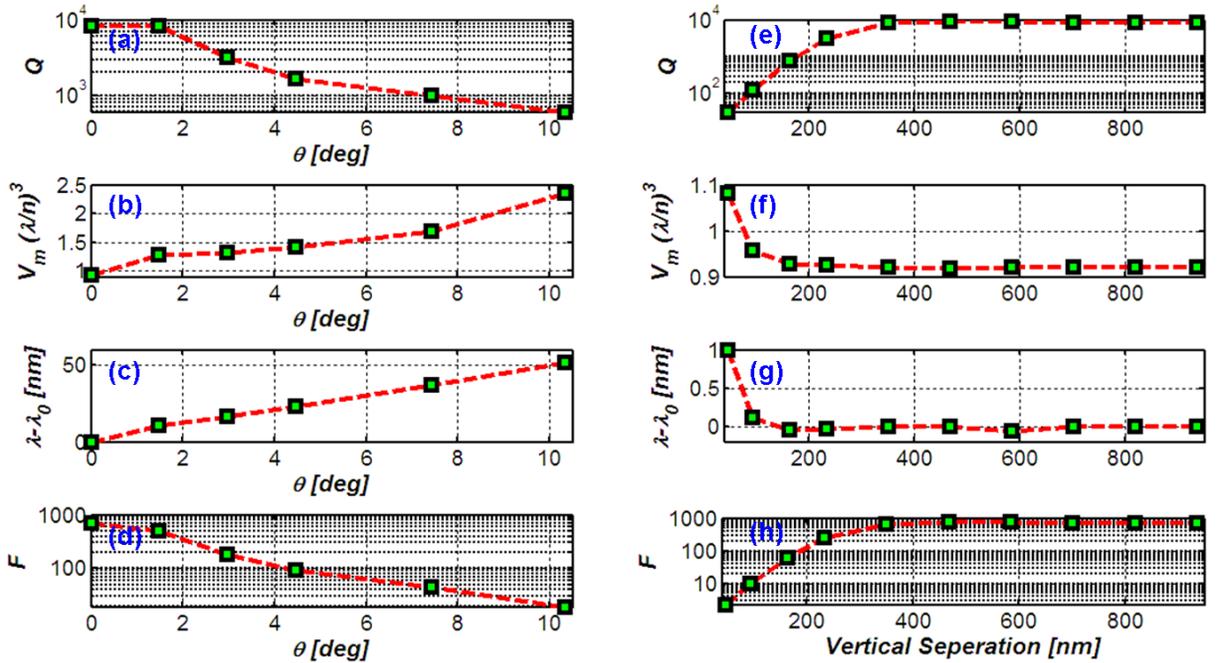

Fig. 4 − The hole wall inclination and substrate proximity effects: (a-d) The mode quality factor (*Q*), the mode volume ($V_m$), the wavelength mode detuning ($\lambda-\lambda_0$) and the Purcell coefficient *F* are given as the function of the hole angle. (e-h) The $Q, V_m, \lambda-\lambda_0,$ and *F* are given as a function of the vertical separation between the membrane and the substrate.

2) Vertical mode confinement: Slab photonic crystals are generally modeled with an infinite air-cladding beneath and above the membrane. In the present sample, due to the membrane fabrication technology, the air gap separation between the membrane and the substrate is about *50nm*. Due to membrane bending this value can locally be even *smaller*. This proximity to the substrate produces leakage of the mode. The influence of the substrate-membrane separation is shown in Figs. 4(e-h). As one can noticed for a *~50nm* separation, *Q=32,* which is *~260* times lower than the original value, while the $V_m$ and the mode frequency are only slightly modified. In order to achieve $Q\sim10^4$, a vertical separation of *~500nm* between the membrane and the substrate is required. To provide this separation with present technology, the membrane should be cut off from the diamond substrate, removed, and relocated onto a pre-prepared host well on top of a foreign substrate (Silicon in the present case).

3) Membrane thickness: In photonic crystals realized in a 2D-slab, the thickness of the slab, providing guidance of photons in the vertical dimension, must be close to the lattice constant, in order to maintain a single mode condition. If the membrane (the slab) is too thick (*h>300nm* in the present case) the cavity quality factor deteriorates, as is discussed bellow. Moreover, variations in the slab thickness may result in a shift of the cavity mode frequency, and its wavelength may fall outside the diamond photoluminescence region of. For example, an increase in membrane thickness by *~50nm* increases the cavity wavelength by *~20nm* and deteriorates *Q* by *60%*.

Table 2 illustrates the sensitivity of the expected *Q* to the variation in design parameters. Listed are the variation of the values quoted to produce a *20%* reduction in the value of *Q*. The most stringent requirements are imposed on the departure from verticality of the walls and on variations in the radius of the PC holes. Note that these parameters are in reality linked, since vertical walls require an over-milling strategy which reduces control over the radius and modifies the membrane thickness. In addition, the hole inclination is influenced by the curvature of the membrane due to the air-gap beneath it during the milling.

Table 2 − Geometrical changes producing *20%* deterioration in the cavity *Q*

| Design Parameter | Value for *0.8×Q* |
|---|---|
| Wall Angle $\theta$ | 1.95° |
| Hole radius $\Delta r$ | 5nm |
| Hole pitch $\Delta a$ | 18nm |
| Membrane Thickness $\Delta h$ | 35nm |

Since the above discussed requirements in the fabrication accuracy are beyond our tool capabilities we have designed and fabricated a series of 25 PC cavities that form a *5×5* matrix, where the rows thickness is varying from *200nm* to *240nm*, while in each column the PC radius is varying from *60nm* to *80nm* (see Fig. 5a). The detailed close-up of a single cavity is shown in Fig. 5b.

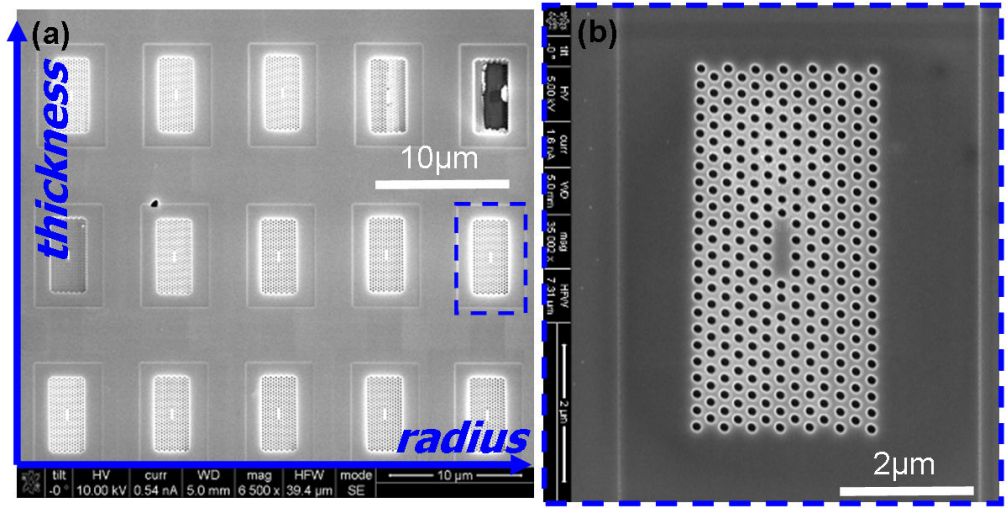

Fig. 5 − (a) Part of the PC matrix with the variation in the thickness and hole radii. (b) Particular cavity close-up view.

## 4. Detailed Fabrication Processing

In this section we review all essential steps towards PC fabrication in diamond. We describe in details membrane formation, PC patterning by Focused Ion Beam and the membrane relocation technique.

4.1 Membrane Fabrication

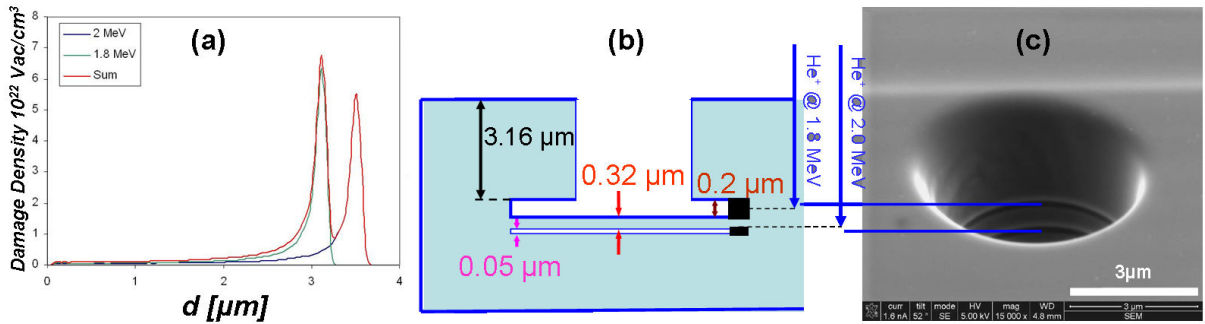

Fig. 6 − Diamond Membrane Formation: (a) Monte-Carle simulation of double-implantation induced damages. (b) Schematic diagram of implanted layers. (c) Two membrane SEM view.

As mentioned above, the formation of a diamond slab (a membrane) separated from the substrate is a pre-requirement for the fabrication of the PC cavity with a high $Q$-factor. The formation of the membrane is based on the well-known phenomenon of the transformation of heavily ion implantation damaged diamond, (in excess of some critical damage level) to graphite during annealing at high temperature [31]. In the present procedure the membrane is formed by first creating a sub-surface sacrificial layer that is subsequently etched. Ion implantation is used to form highly damaged layers at different depths and thicknesses depending on implantation conditions. To obtain the required membrane thickness two He+ ion implantations at energies of 1.8MeV and 2MeV are applied (see Fig. 6a). In the low damage density regions, below the critical damage density, thermal annealing induces diamond lattice recovery, while the regions with damage density in excess of critical value become fully graphitized [31]-[33]. The graphitized layers are wet etched (as described below) via FIB milled trenches which enable the penetration of the etchant, resulting in two free-standing membranes *3.16* and *0.32μ* thick (Fig. 6(b,c)). Removal of the upper *3.16μm* membrane allows access and patterning of the PC structure on the thin film. In Fig. 6b ,a schematic of the expected thickness of each layer is shown.

4.2 Photonic Crystal FIB Patterning

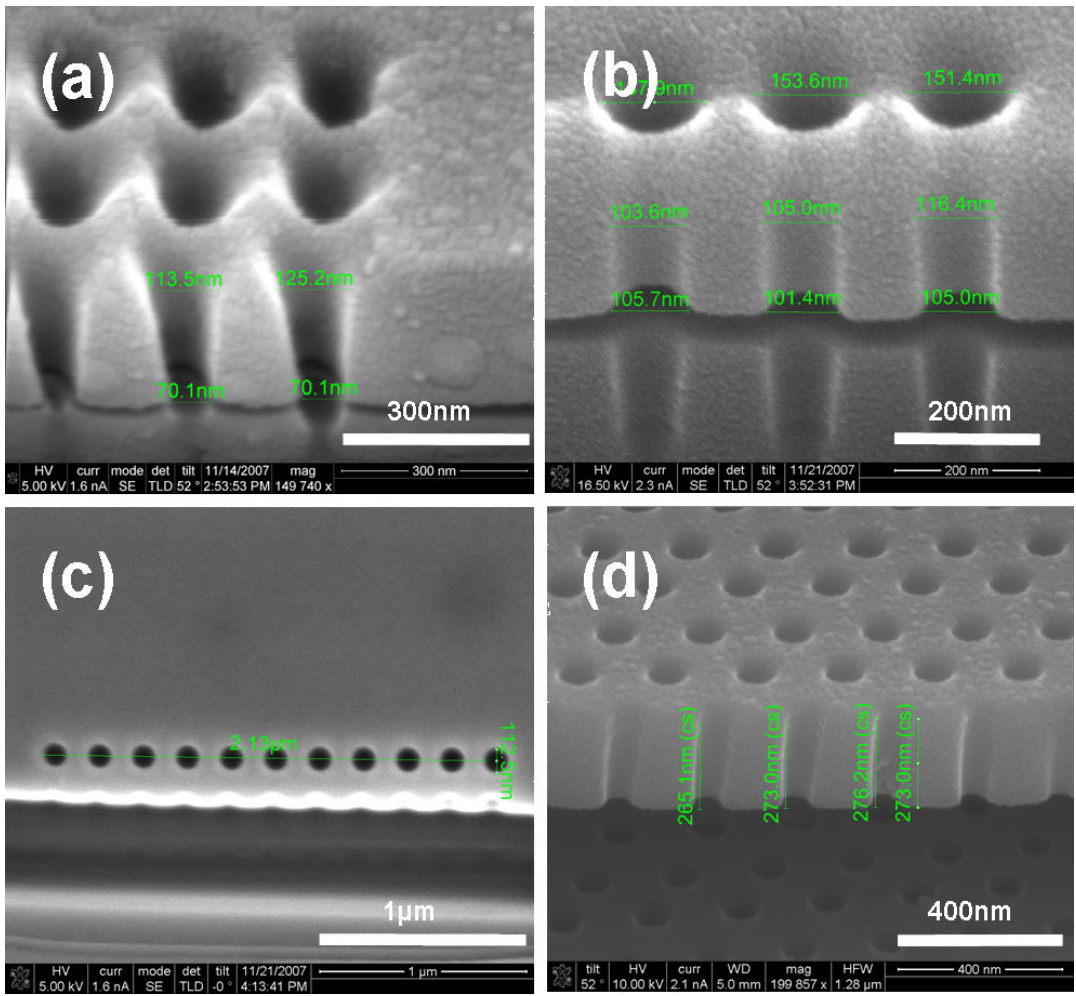

Fig. 7 − High resolution SEM of the PC obtained by FIB milling. Hole parameters characterization: (a) Non-vertical holes due to low air-gap. (b) Holes with high verticality. (c) Lateral hole profile (the same holes as in (b)). (d) Modified thickness caused by overmilling.

The photonic crystal is composed by a triangular array of holes patterned onto the thin membrane by FIB milling. To avoid charging and electromagnetic distortion during the FIB processing, the membrane is covered with an evaporated ~40nm thick chromium conductive layer. As an example of the present process, a PC with hole radius *r=57nm* and pitch of *212nm* was fabricated. Using these parameters, that are expected to fit [24] with mode wavelength of *637nm* ($NV^-$ center wavelength), a moderate-*Q* cavity was designed (Table-1). FIB patterning of the PC holes (milling) is achieved by using a $Ga^+$ ion focused beam with a current of *28pA* at an accelerating voltage of *30kV*. Following this step the sample was annealed at 700°C and acid cleaned as described below. Since diamond is etched at about half the rate of Si, the present process represents a ×5 over etching, as compared to that expected for Si (which is tabulated) in order to obtain profiles as vertical as possible.

The PC-hole profile (deviation from vertical walls) during FIB milling is also affected by the air-gap under the thin membrane. The air-gap is nominally *l=50nm*, however, due to membrane bending the gap can vary considerably. We have measured the wall inclination angle to the vertical, $\theta$, using the SEM images, to vary in the range *1°< $\theta$<5.3°*, the higher angle corresponds to the lower-gap region due to a slow milling product disposal (Fig. 7a,b). The thickness of the gap is difficult to control and it is only roughly known after the milling process. In Fig. 7(c) we show the top view of PC hole cross-section presenting an excellent circularity in addition to the vertical wall profile. The hole overmilling modifies the membrane thickness in the PC area due to the Gaussian profile of the ion beam at the upper surface, and the merging of neighboring holes. As a result, the thickness in the area of the PC is *~270nm* while the thickness of the membrane without the PC patterning is *330nm* (see Fig7c). Therefore, the pattern production makes the membrane *~60nm* thinner. After the milling, the sample is wet-etched to remove the chromium with a standard etchant. Re-deposited milling products are mostly removed in a boiling $HClO_4/HNO_3/H_2SO_4$ (1:1:1) acid-mix. After etching, the sample was carefully inspected under a microscope, and no traces of residue were observed. It should, however be noted that some shallowly implanted Ga is expected to be present at and near the FIB treated areas.

4.3 Membrane Relocation

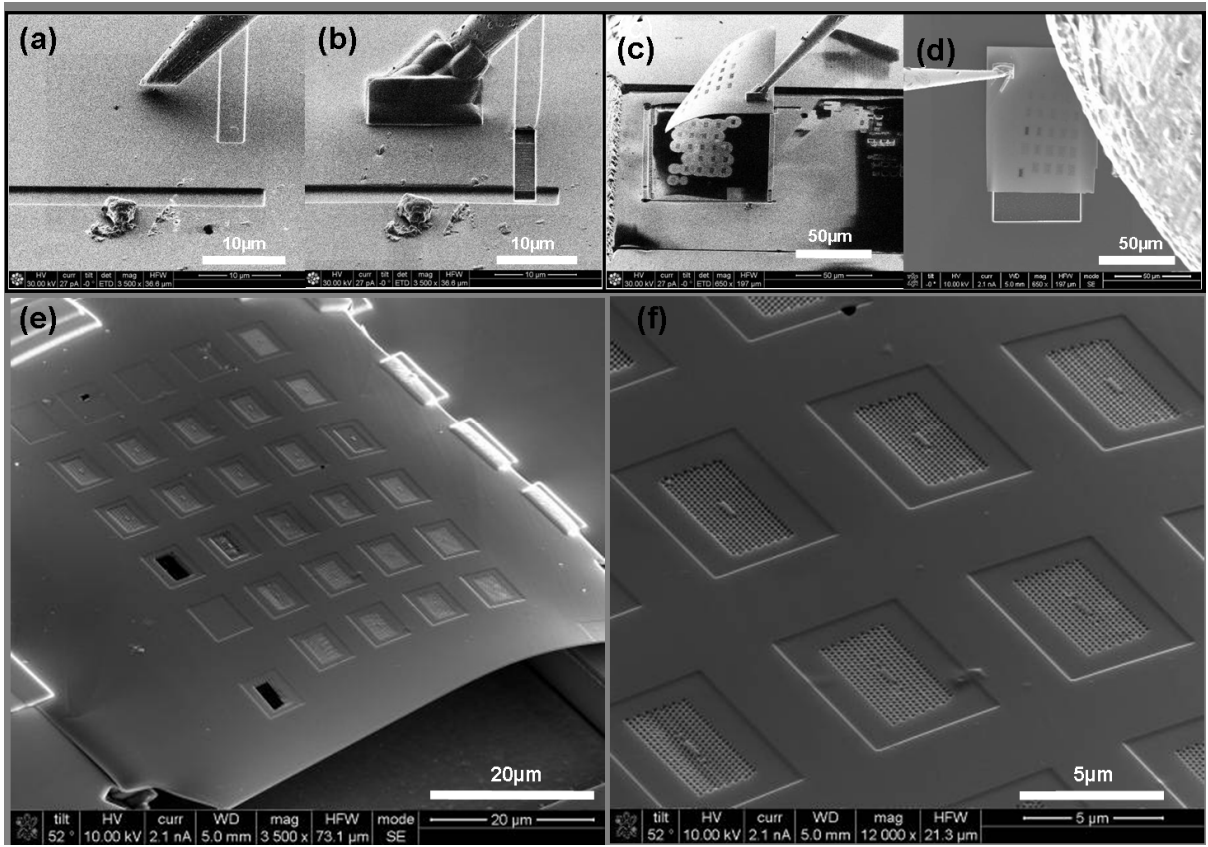

Fig. 8 − The relocation of PC. Each picture presents SEM/FIB micrograph displaying the up/front views. (a) The manipulator needle positioning on the sample edge. (b) Pt welding of the needle to the membrane. (c) The lift-off of the membrane after final cut-off of the edges. (d) The membrane positioning over the well. (e,f) Relocated PC Matrix: The side view and the close-up on several cavities.

In order to provide a larger air-gap below the membrane (for improvement of the vertical optical confinement) the membrane with its PC's was relocated within the FIB machine and placed onto a pre-prepared well in silicon several micrometers deep. The relocation process is demonstrated in Fig. 8. First, the membrane is partially cut-off. Then, a micromanipulator needle is attached to an edge of the membrane (already partly cut off and cleaned from re-deposits), and is welded to it by platinum deposition. Then, the membrane is fully cut free by the FIB and is lifted-off from its diamond host and relocated onto the silicon well. The membrane is fixed to the silicon in the overlap regions along its edges by a platinum deposition. The manipulator is then cut-off from the membrane. Note that during the relocation the diamond membrane is not covered with the conductive layer, as is generally the case during the Transmission-Electron-Microscopy (TEM) sample preparation [36]-[38], thus the relocation is performed on pure dielectric material with partial charge removal by the manipulator. As a result, the free standing membrane with the PC series in it is located on the pre-prepared well (see Fig. 8e,f) and it can be subjected to optical characterization.

**5. Optical Characterization**

The set of thus realized PC devices was characterized by a confocal-micro-photoluminescence set-up, in which the cavity mode is excited by light from a green laser ($\lambda=532nm$) via a microscope objective of numerical aperture $NA=0.9$. The photoluminescence is separated from the excitation by a dichroic mirror and a visible-window filter at the entrance of the monochromator. The spectrum is measured at room temperature by a $0.75m$ monochromator with a grating of 1200 $mm^{-1}$ and a nitrogen-cooled CCD camera. The results are shown in Fig. 9.

The characteristic spectrum of unprocessed Type-Ib diamond is shown in the inset of Fig. 9a. It contains both $NV^0$ and $NV^-$ peaks and it serves as a reference in the following discussion. To evaluate the properties of the relocated membrane, the spectrum of an unpatterned upper thick membrane, cut-off and positioned onto the Si substrate was taken. This membrane is most likely affected by implantation of $He^+$ (for the formation of the membrane) and by $Ga^+$ (related to FIB process). In addition, its upper side was covered with a thin layer of chromium to avoid charging during FIB manipulation. The measurement is performed by the confocal set up. By moving the confocal pin-hole (z-axis) spectra sampling in various membrane depth-slices between the upper and the lower surfaces is obtained. The resulted slice-spectra are shown in Fig. 9a. In the vicinity of both surfaces the NV$^-$ peak is reduced and apparently mostly transformed into NV$^0$ [35]. On the upper surface this transformation might be related to either Ga$^+$ implantation or chromium cover, while on the lower one it may originate from He$^+$ implantation generated defects, and/or scattered Ga$^+$ ions during FIB processing. In addition, there are 4 clear spectral peaks corresponding to Fabry-Perot oscillation modes formed by the membrane upper and lower interfaces ($\lambda=658, 693, 727, 763nm$), with an average peak distance of $35nm$. By relating the Fabry-Perot Free Spectral Range ($FSR$)=$35nm$ to the membrane thickness $h$ ($FSR=\lambda^2/(2nh)$, where $n$ is the refractive index of diamond, $n=2.4$)) an estimate for the membrane thickness of $h=3.05\mu m$ is obtained. This is very close to the designed membrane thickness ($h=3.16\mu m$). Thus, the thick membrane has sufficient optical quality to act as an interferometer.

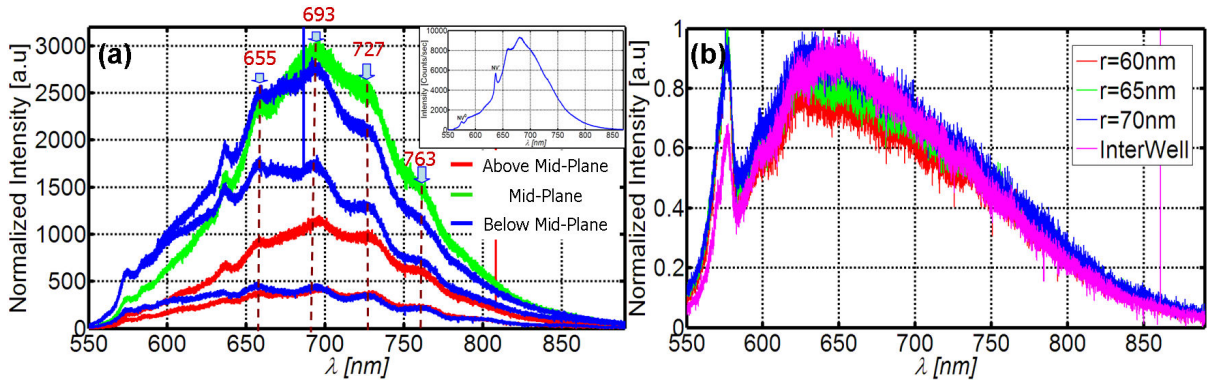

Fig. 9 − The photoluminescence measurements. (a) The measurement of the Cr covered relocated thick membrane $h=3.16\mu$. The spectrum is taken at various thickness planes. Red defines the color above the mid membrane, blue below the mid and green at the middle. In the inset PL spectrum of Type-Ib diamond. (b) Characteristic spectrum of PC cavities of similar thickness and various hole

radii of the PC. The magenta spectrum is taken in the inter-well region. No sign of PC resonance is observed. The wide peak at $\lambda=576.5nm$ is attributed to $NV^0$.

In Fig. 9b the characteristic spectra of the relocated PC cavities are shown for different hole radii. The areas inside and outside of the PC cavity region are presented. The outside spectrum is taken in the inter-well region. As one can observe, the $NV^-$ characteristic peak at *637nm* is hardly visible after processing and relocation of the thin membrane. Instead, a very high peak at *576.5 nm*, which is shifted by *1.5nm* with respect to that in the unprocessed diamond is observed (see Fig. 1a). Although this might well be attributed to $NV^0$, there is also the possibility that it may be a feature due to the presence of Ga. This peak is by *30%* higher in the cavity center than outside the cavity well. However, the shapes of the spectra are similar in both regions. Spectral features characteristic of PC cavity resonances were not observed in this measurements. The absence of observable cavity resonances may be an indication of degraded optical properties of the membrane. Since $1/Q=1/Q_{cav}+1/Q_{loss}$ ($Q_{loss}$ being process related losses or scattering effects), one may conclude that the $Q_{loss}$ is too low to sustain the cavity confinement. This may be attributed to poor optical properties of the diamond/air interfaces possibly caused by either $Ga^+$ implantation during the patterning or to residual $He^+$ implantation in the membrane formation processing The disappearance of the $NV^-$ can be explained by residual damage due to the ion implantation or due to the hole processing causing quenching of $NV^-$ [35]. This is supported by the characterization of the edges in the thick membrane in which the most pronounced spectrum originates from the middle of the sample, i.e. being furthest away from the edges (Fig. 9a).

The fact that the thick diamond membrane whose upper edge has been less exposed to implantation damage does exhibit the expected optical properties, while the thin one does not, may indicate that that optical losses are mainly related to the response of the diamond to unremoved residual implantation damage, however, other fabrication related damage, such as the damage caused by milling and losses due to absorption, scattering, surface texturing, changes in material properties can not be ruled out. Hence removal of the residual damaged part of the slab still remains a major challenge of the present fabrication approach. Recent results suggest that H plasma treatment is useful in removing most of the damaged material, leaving behind a nearly pristine diamond surface [39].

## 6. Summary

A comprehensive study of the parameter space for the realization of photonic crystals in a thin diamond membrane has been presented. A complete sequence of processing steps for implementation of photonic crystal nano-cavities on monocrystalline diamond were described. The processing procedures (i.e. hole-profile, substrate separation, membrane thickness and hole radius) and the required tolerances were presented. The fabrication approach is based on membrane formation by Ion Implantation, Focused Ion Beam patterning of 2D-slab photonic crystal cavities, and micro-probe handling of the patterned slab for relocation. These process steps are likely to become part of a toolbox useful in the future implementation of photonic crystal architecture for diamond-based QED. As has been shown from the PL measurements only the thick diamond membrane (not exposed to ion beam processing) exhibits the optical quality required for the realization of photonic devices. In the present approach, the thin membrane fabrication process combined with the photonic crystal patterning determines the optical properties for this application. At the moment, the expected spectral features of the fabricated devices are not observed, thus, understanding of the factors limiting the optical properties of the processed surfaces is needed. Further development of the processing procedure or finding alternative approaches to realize nano-cavities in diamond are needed to satisfy the requirements of photonic applications and quantum information implementation in diamond.

**Acknowledgements**

RK, FJ and JW acknowledge partial support of European Union NanoEngineered ERANET. The support of the, Russell Berrie Nanotechnology Institute (RBNI) at the Technion is acknowledged. B.M. acknowledge the support of KAMEA program (Israel). OB acknowledges partial support by DFG (grant BE2224/9) and BMBF (KEPHOSI**).** Discussion with A. D. Greentree and S. Tomljenovic-Hanic are acknowledged. The support of the Australian Research council through the linkage international scheme is gratefully acknowledged.